\documentclass[12pt]{article}

\usepackage{mathrsfs}
\usepackage[T1]{fontenc}
\usepackage{mathpazo}
\usepackage{setspace}
\usepackage{amsfonts}
\usepackage{amssymb}
\usepackage{amsmath}
\usepackage{esint}                   
\usepackage{epsfig}
\usepackage{latexsym}
\usepackage{color}
\usepackage{graphicx}
\usepackage{hyperref}
\usepackage{nicefrac}
\usepackage[latin1]{inputenc}
\usepackage{pstricks}
\usepackage{slashed}
\usepackage{multirow}
\usepackage{ulem}
\usepackage{comment}
\usepackage[nosort]{cite}
\usepackage{datetime}
\usepackage{tikz-cd}
\usepackage{float}

\def\hybrid{\topmargin -30pt    \oddsidemargin 0pt 
        \headheight 0pt \headsep 0pt
        \textwidth 6.25in       
        \textheight 9.5in       
        \marginparwidth .875in
        \parskip 5pt plus 1pt   \jot = 1.5ex}

\hybrid

\def\baselinestretch{1.2}

\catcode`\@=11

\def\marginnote#1{}
\newcount\hour
\newcount\minute
\newtoks\amorpm
\hour=\time\divide\hour by60
\minute=\time{\multiply\hour by60 \global\advance\minute by-\hour}
\edef\standardtime{{\ifnum\hour<12 \global\amorpm={am}%
        \else\global\amorpm={pm}\advance\hour by-12 \fi
        \ifnum\hour=0 \hour=12 \fi
        \number\hour:\ifnum\minute<10 0\fi\number\minute\the\amorpm}}
\edef\militarytime{\number\hour:\ifnum\minute<10 0\fi\number\minute}

\def\draftlabel#1{{\@bsphack\if@filesw {\let\thepage\relax
   \xdef\@gtempa{\write\@auxout{\string
      \newlabel{#1}{{\@currentlabel}{\thepage}}}}}\@gtempa
   \if@nobreak \ifvmode\nobreak\fi\fi\fi\@esphack}
        \gdef\@eqnlabel{#1}}
\def\@eqnlabel{}
\def\@vacuum{}
\def\draftmarginnote#1{\marginpar{\raggedright\scriptsize\tt#1}}

\def\draft{\oddsidemargin -.5truein
        \def\@oddfoot{\sl preliminary draft \hfil
        \rm\thepage\hfil\sl\today\quad\militarytime}
        \let\@evenfoot\@oddfoot \overfullrule 3pt
        \let\label=\draftlabel
        \let\marginnote=\draftmarginnote
   \def\@eqnnum{(\theequation)\rlap{\kern\marginparsep\tt\@eqnlabel}%
\global\let\@eqnlabel\@vacuum}  }

\def\draft2{
        \def\@oddfoot{\sl preliminary draft \hfil
        \rm\thepage\hfil\sl\today\quad\militarytime}
        \let\@evenfoot\@oddfoot \overfullrule 3pt
        \let\marginnote=\draftmarginnote
   \def\@eqnnum{(\theequation)\rlap{\kern\marginparsep\tt\@eqnlabel}%
\global\let\@eqnlabel\@vacuum}  }

\def\preprint{\twocolumn\sloppy\flushbottom\parindent 2em
        \leftmargini 2em\leftmarginv .5em\leftmarginvi .5em
        \oddsidemargin -.5in    \evensidemargin -.5in
        \columnsep .4in \footheight 0pt
        \textwidth 10.in        \topmargin  -.4in
        \headheight 12pt \topskip .4in
        \textheight 6.9in \footskip 0pt
        \def\@oddhead{\thepage\hfil\addtocounter{page}{1}\thepage}
        \let\@evenhead\@oddhead \def\@oddfoot{} \def\@evenfoot{} }

\def\numberbysection{\@addtoreset{equation}{section}
        \def\theequation{\thesection.\arabic{equation}}}

\def\underline#1{\relax\ifmmode\@@underline#1\else
        $\@@underline{\hbox{#1}}$\relax\fi}

\def\titlepage{\@restonecolfalse\if@twocolumn\@restonecoltrue\onecolumn
     \else \newpage \fi \thispagestyle{empty}\c@page\z@
        \def\thefootnote{\fnsymbol{footnote}} }

\def\endtitlepage{\if@restonecol\twocolumn \else \newpage \fi
        \def\thefootnote{\arabic{footnote}}
        \setcounter{footnote}{0}}  

\catcode`@=12
\relax

\def\figcap{\section*{Figure Captions\markboth
        {FIGURECAPTIONS}{FIGURECAPTIONS}}\list
        {Figure \arabic{enumi}:\hfill}{\settowidth\labelwidth{Figure
999:}
        \leftmargin\labelwidth
        \advance\leftmargin\labelsep\usecounter{enumi}}}
 \relax
\def\tablecap{\section*{Table Captions\markboth
        {TABLECAPTIONS}{TABLECAPTIONS}}\list
        {Table \arabic{enumi}:\hfill}{\settowidth\labelwidth{Table
999:}
        \leftmargin\labelwidth
        \advance\leftmargin\labelsep\usecounter{enumi}}}
 \relax
\def\reflist{\section*{References\markboth
        {REFLIST}{REFLIST}}\list
        {[\arabic{enumi}]\hfill}{\settowidth\labelwidth{[999]}
        \leftmargin\labelwidth
        \advance\leftmargin\labelsep\usecounter{enumi}}}
 \relax
\makeatletter
\newcounter{pubctr}
\def\publist{\@ifnextchar[{\@publist}{\@@publist}}
\def\@publist[#1]{\list
        {[\arabic{pubctr}]\hfill}{\settowidth\labelwidth{[999]}
        \leftmargin\labelwidth
        \advance\leftmargin\labelsep
        \@nmbrlisttrue\def\@listctr{pubctr}
        \setcounter{pubctr}{#1}\addtocounter{pubctr}{-1}}}
\def\@@publist{\list
        {[\arabic{pubctr}]\hfill}{\settowidth\labelwidth{[999]}
        \leftmargin\labelwidth
        \advance\leftmargin\labelsep
        \@nmbrlisttrue\def\@listctr{pubctr}}}
 \relax
\makeatother

\def\be{\begin{equation}}
\def\ee{\end{equation}}
\def\ba{\begin{eqnarray}}
\def\ea{\end{eqnarray}}

\def\del{\partial}

\def\k{\kappa}
\def\r{\rho}

\def\b{\beta}

\def\G{\Gamma}

\def\D{\Delta}

\def\th{\theta}

\def\m{\mu}
\def\n{\nu}

\def\Om{\Omega}
\def\l{\lambda}
\def\L{\Lambda}
\def\s{\sigma}

\def\no{\noindent}

\def\IR{\relax{\rm I\kern-.18em R}}

\def\inv{^{\raise.0ex\hbox{${\scriptscriptstyle -}$}\kern-.05em 1}}

\def \ov {\over}

\begin{document}


\renewcommand{\theequation}{\thesection.\arabic{equation}}
\csname @addtoreset\endcsname{equation}{section}

\begin{titlepage}
\begin{center}

\renewcommand*{\thefootnote}{\arabic{footnote}}

\hfill  HU-EP-21/30
\phantom{xx}
\vskip 0.5in

{\large {\bf Kerr--Schild perturbations of coset CFTs\\
\vskip 0.03 cm
as scale invariant integrable $\sigma$-models}}

\vskip 0.5in

{\bf Georgios Itsios},\footnote{E-mail:~gitsios@phys.uoa.gr, \, georgios.itsios@physik.hu-berlin.de}$^{a,b}$\hskip .2cm
{\bf Konstantinos Sfetsos}\footnote{E-mail:~ksfetsos@phys.uoa.gr}$^{a}$\hskip .15cm and \
{\bf Konstantinos Siampos}\footnote{E-mail:~konstantinos.siampos@phys.uoa.gr}$^{a}$

\vskip 0.1in

$^a$Department of Nuclear and Particle Physics, \\
Faculty of Physics, National and Kapodistrian University of Athens, \\
Athens 15784, Greece \\
$^b$Institut f\"{u}r Physik, Humboldt-Universit\"{a}t zu Berlin,\\
IRIS Geb\"{a}ude, Zum Gro{\ss}en Windkanal 2, 12489 Berlin, Germany
\vskip .3 cm


\vskip .2in

\end{center}

\vskip .4in

\centerline{\bf Abstract}
\no
 Kerr--Schild perturbations in General Relativity provide a fruitful way of constructing new exact solutions starting from known ones,
elucidating also the structure of the spacetimes. We initiate such a study in the context of string theory and supergravity.
Specifically, we explicitly construct Kerr--Schild perturbations of coset CFTs based on low dimensionality orthogonal groups.
We show that these give rise to scale, but not Weyl, invariant integrable $\sigma$-models. We explicitly demonstrate that these models can also be derived from
a particular limiting procedure of $\lambda$-deformed coset CFTs based on non-compact groups. The target space of the simplest $\sigma$-model 
describes a two-dimensional scale invariant black hole for which we also provide two different embeddings to type-II supergravity.

\vfill

\end{titlepage}
\vfill
\eject



\def\baselinestretch{1.2}
\baselineskip 20 pt

\newcommand{\eqn}[1]{(\ref{#1})}

\tableofcontents

\section{Introduction}

In General Relativity the Kerr--Schild perturbation and its generalizations provide a rather powerful technique in generating new distinct solutions from known ones~\cite{KSpaper,GGPaper,Xanthopoulos1,Taub,Dereli:1986cm,Xanthopoulos2} (for a review see ch. 32 in~\cite{Stephani}). Most of the exact solutions to Einstein's equations can be obtained by using this 
procedure, including the Schwarzschild, Reissner--Nordstr\"om, Kerr and Kerr--Neumann black hole solutions by perturbing a flat/dS/AdS metric. 

The Kerr--Schild deformation is defined in terms of a null geodesic vector  $\ell^\mu$  as 
\be
\label{gen.KS.metric}
g_{\mu\nu}=g^{(0)}_{\mu\nu}+V\ell_\mu\ell_\nu\,,\quad
g^{(0)}_{\mu\nu}\ell^\mu\ell^\nu=0\,,\quad \ell^\mu\nabla^{(0)}_\mu\ell_\nu=0\,,
\ee
where $V$ is the Kerr--Schild potential and the covariant derivatives are constructed using the Levi--Civita connection 
of the $g^{(0)}_{\mu\nu}$. We note that the same conditions also hold for the metric $g_{\mu\nu}$. 
The above procedure was initiated for Minkowski space-time $\eta_{\m\n}$ in~\cite{KSpaper,GGPaper,Xanthopoulos1} and later
extended for general metrics 
$g^{(0)}_{\m\n}$ in~\cite{Taub,Dereli:1986cm,Xanthopoulos2}. 
The advantage of this special type of rank-1 perturbations to the metric is that the linearized form of the Einstein equations 
is also exact~\cite{KSpaper,Taub, Dereli:1986cm}. In addition, the demand for a null vector $\ell^\m$ to exist restricts the Kerr--Schild formulation to spaces with Minkowskian signature.

The simplest example is the static Ricci flat Schwarzschild black hole which is obtained by taking the Minkowski space as the original seed solution.
Specifically, we have that
\be
\text{d}s_{(0)}^2=-\text{d}t^2+\text{d}r^2+r^2(\text{d}\theta^2+\sin^2\theta\text{d}\phi^2)\,,\quad V=\frac{2GM}{r}\,,\quad
\ell=-\text{d}t+\text{d}r
\ee
and we may also include rotation, charges and cosmological constant. 
Much to our surprise the Kerr--Schild procedure has not been applied in a string theoretical context.  We undertake this task by
focusing on some low dimensionality CFTs with an interesting spacetime interpretation.
We will provide exact results in the perturbative Kerr--Schild term. In order to emphasize that aspect
we will freely call such perturbations as deformations as well.

As a concrete example we work out such deformations for the $\nicefrac{SL(2,\mathbb{R})_{-k}}{SO(2)}$ coset CFT~\cite{wittenbh}.
The resulting model turns out to be classically integrable and its charges are in involution. In addition, its $\b$-function is vanishing and the corresponding diffeomorphism
cannot be written in terms of a scalar (dilaton). This type of model is scale invariant rather than conformal~\cite{Hull:1985rc}. In fact, it 
becomes Weyl invariant after a correlated asymptotic limit is taken.
Interestingly, we found that this type of deformations could be also obtained from a particular limit in the corresponding $\l$-deformed $\s$-models, 
constructed in~\cite{Sfetsos:2013wia}. Moreover, this model can be embedded in type IIA/IIB SUGRA using appropriately the results of~\cite{Sfetsos:2014cea}.
Finally, we extend our results in a higher dimensional model with a non-compact target spacetime, such as the  $\nicefrac{SO(2,2)_{-k}}{SO(1,2)_{-k}}$ coset CFT~\cite{Bars:1992ti}.
The derived $\s$-model is again scale invariant and classically integrable with the corresponding conserved charges in involution.

This work is structured as follows: In Section \ref{Sec:2}, we consider the $\nicefrac{SL(2,\mathbb{R})_{-k}}{SO(2)}$ coset CFT~\cite{wittenbh}
which we perturb with a Kerr--Schild deformation and we obtain a classically integrable scale invariant $\s$-model.  
In addition, we derive the scale invariant model as an asymptotic-like limit of the $\l$-deformed $\nicefrac{SL(2,\mathbb{R})_{-k}}{SO(2)}$ model and we 
study its embedding in type IIA/IIB-supergravity. In Section~\ref{Sec:3}, we consider the $\l$-deformed $\nicefrac{SO(2,2)_{-k}}{SO(1,2)_{-k}}$ $\s$-model
and we apply an analogue asymptotic-like limit yielding a Kerr--Schild deformation of the underlying coset CFT, in its asymptotic region.
This deformation corresponds to a classically integrable and scale invariant $\s$-model as well. 
Finally, Appendix~\ref{appendix.KS.gen} provides 
an exposure of the Kerr--Schild deformation, some relevant results and  the presentation, as an important  example, of the renowned Kerr solution
in the presence of a cosmological constant.

\section{Kerr--Schild deformation of a 2-dim target space CFT}
\label{Sec:2}

\no
In this section we will consider a Kerr--Schild deformation of the $\nicefrac{SL(2,\mathbb{R})_{-k}}{SO(2)}$ coset CFT~\cite{wittenbh}
and we will obtain a scale invariant $\s$-model. This model can be also derived from a correlated asymptotic type limit 
of the $\l$-deformed $\s$-model constructed in~\cite{Sfetsos:2013wia}.

\subsection{Scale invariance}

The simplest non-trivial example arises by considering the
$\nicefrac{SL(2,\mathbb{R})_{-k}}{SO(2)}$ coset CFT with a non-compact two-dimensional target space~\cite{wittenbh}.
The corresponding $\s$-model action is
\be
\label{2dCFT}
S_{(0)}=\frac{k}{\pi}\int\text{d}^2\sigma\big(\del_+\r\del_-\r  - \coth^2 \r \, \del_+\tau\del_-\tau\big)\, .
\ee
In addition, there is the dilaton field 
\be
\label{2d.dilaton}
\Phi=-\ln\sinh\rho\,.
\ee
The above spacetime is obtained in the vector gauging of $SO(2)$ in $SL(2,\mathbb{R})$ and represents the region of a black hole
 beyond the singularity \cite{wittenbh}.
The above target space in the action \eqref{2dCFT} is quite simple and allows for the generic null and geodesic vector
\be
\label{null.2d}
\ell=\text{d}\tau+\tanh\rho\,\text{d}\rho\,.
\ee
The function $V(\tau,\rho)$ appearing in the analogue of \eqref{gen.KS.metric}
\be
\label{g0gll}
g_{\mu\nu}=g^{(0)}_{\mu\nu}+\l V\ell_\mu\ell_\nu\,,\quad X^\m=(\tau,\rho)\,,
\ee
remains still undetermined and will be fixed by demanding the one-loop renormalizability of the $\s$-model.
As we will see all the renormalization effects will be absorbed  in the multiplicative real parameter $\l$ appearing in \eqref{g0gll}.

\no
 For a $\s$-model with a general metric
 \be
\label{gen2d}
S=\frac{1}{2\pi}\int\text{d}^2\s\,  g_{\m\n}\del_+X^\m\del_-X^\n\, ,
\ee
 the RG flows at one-loop order are given by~\cite{Ecker,Friedan:1980jf}
\begin{equation}
\label{RG.oneloop}
 \frac{\text{d}  g_{\m\n}}{\text{d}t} =  R_{\m\n} +  \nabla_\mu \xi_\nu+ \nabla_\nu\xi_\mu\,,
\end{equation}
where $t=\ln\mu^2$ and $\mu$ is the energy scale. At present, we will consider a diffeomorphism of the type (it turns out to be the only consistent choice)
\be
\label{KS.diff}
\xi= g^{\m\n}\del_\n\Phi\del_\m+\eta\,,\quad \eta=-\frac{c}{2k}\del_\tau\,,
\ee
with $c$ being an arbitrary parameter. In the second term, $\eta$ cannot be written in terms of a new scalar since
\be
\label{qshsoq1}
\del_\m\eta_\n-\del_\n\eta_\m\neq0\ .
\ee
 It turns out that its Lie derivative satisfies
\be
\label{qshsoq11}
\nabla_\mu\eta_\nu+\nabla_\nu\eta_\mu=-\l\frac{c}{k}V\ell_\mu\ell_\nu\,,
\ee
demonstrating that $\eta$ is a Killing vector at the conformal point $\l=0$, as it can be clearly seen from \eqn{2dCFT}.
Inserting \eqref{2dCFT}, \eqref{2d.dilaton}, \eqref{null.2d} and \eqref{KS.diff} into \eqref{RG.oneloop} we find that 
 consistency requires that
\be
\label{V.2d}
V(\tau,\rho)=2k\left(\text{e}^\tau\cosh\rho\right)^{\frac{2f(\l)}{(1-c)\l}}(c_1+\coth^2\rho)\,,
\ee
where $c_1$ is an integration constant.
In addition, the function $f(\l)$ parameterizes the one-loop RG flow of the parameter $\l$ as
\be
\label{RG.KS}
\frac{\text{d}\l}{\text{d}t}=\frac{f(\l)}{k}\,
\ee
and $k$ does not flow.
To constrain the parameter $c_1$ we will demand that the function $V(\tau,\rho)$  stays finite
as $\rho\gg1$, similarly to the coset CFT $\s$-model action \eqref{2dCFT} and moreover it remains non-trivial.
This demand uniquely sets $c_1=-1$\footnote{\label{Weyl.foot}This is in agreement with
 the Weyl anomaly coefficient (playing a r\^ole of a c-function)
\begin{equation*}
W=2-3(R+4\nabla^2\Phi-4(\nabla\Phi)^2)=2+\frac{6}{k}-\frac3k\l(1+c_1)\left(\text{e}^\tau\cosh\rho\right)^{\frac{2f(\l)}{\l}}\,,
\end{equation*}
which is independent of $\tau$ and $\rho$ only for $c_1=-1$.} and the function $f(\l)=(1-c)\l$. Using the latter we find that
\be
\label{V.final}
V(\tau,\rho)=2k\,\text{e}^{2\tau}\coth^2\rho
\ee
and the corresponding action reads
\be
\label{action.2dHT}
\begin{split}
S&=  {k\ov \pi}\int\text{d}^2\s\left\{\del_+\rho\del_-\rho-\coth^2\rho\del_+\tau\del_-\tau\right.\\
&\left.+\l\,\text{e}^{2\tau}\coth^2\rho(\tanh\r\,\del_+\rho+\del_+\tau)(\tanh\r\,\del_-\rho+\del_-\tau)\right\}\,,
\end{split}
\ee whereas the scalar $\Phi$ is still given by \eqref{2d.dilaton}. For $\l\neq 0$, we may introduce new coordinates as
\be
u=\left(\sqrt{\l}\text{e}^\tau+\frac{1}{\sqrt{\l}}\text{e}^{-\tau}\right)\cosh\rho\,,\quad v=\sqrt{\l}\text{e}^\tau\cosh\rho\, ,
\ee
so that the $\s$-model action reads
\be
\label{action.2dHTconf}
S=\frac{k}{\pi}\int\text{d}^2\sigma\frac{\del_+u\del_-v}{uv-v^2-1}\, .
\ee
Clearly, the above  change of variables shows that $u$ and $v$ have a restricted range,
which may be extended so that they become global coordinates.

\no
Finally, we comment on the RG flow of $\l$ in \eqref{RG.KS} which, after choosing the function $f(\l)$ as above,
depends explicitly on the parameter $c$
\begin{equation}
\label{c.dep.RG0}
\frac{\text{d}\lambda}{\text{d}t}=(1-c)\frac{\lambda}{k}\,.
\end{equation}
Observe that the parameter $\l$ in \eqref{action.2dHT} can be absorbed by a shift of $\tau$, generated by the $\eta$ diffeomorphism
in \eqn{KS.diff}.
Therefore, on  physical grounds its $\beta$-function cannot explicitly depend on the parameter $c$ appearing in
the latter diffeomorphism. Hence, we should pick $c=1$ corresponding to a {\it scale invariant $\s$-model} rather than a Weyl invariant one.\footnote{ This property does not contradict the results of~\cite{Polchinski.Scale} where it was shown that for any unitary two-dimensional model with a compact target space, scale invariance implies Weyl invariance. The models at hand have non-compact target spaces. 
One may readily verify that any analytic continuation that makes the target space compact gives rise to imaginary, manifestly non-unitary
actions.}
The reason is that the diffeomorphism cannot be written in terms of a scalar \cite{Hull:1985rc} as it was noted in \eqref{qshsoq1}.

\no
The physical interpretation of \eqref{action.2dHTconf} is that of a scale invariant black hole. The corresponding Kruskal--Szekeres diagram is
drawn in Figure~\ref{Kruskal.scale} appearing later in subsection~\ref{xjakqj} and further comments will be presented there.
This will be done in relation and comparison with that for the conformal case
of \cite{wittenbh} and as well as with that corresponding to the spacetime of the $\l$-deformed model for the $\nicefrac{SL(2,\mathbb{R})_{-k}}{SO(2)}$
coset CFT.

\subsection{Integrability}

To prove the classical integrability of the action \eqref{action.2dHT} we need to show that its equations of motion
take the form of a Lax connection. For this purpose, we define
\be
\label{Lax.comp}
\begin{split}
&{\cal L}_\pm^1=\frac{i}{2}\,\text{e}^\tau\,\zeta^{\pm1}\sqrt{\l}\left(\del_\pm\rho+\coth\rho\del_\pm\tau\right)\,,\\
&{\cal L}_\pm^2=\pm\frac12\text{e}^\tau\,\zeta^{\pm1}\sqrt{\l}\left(\del_\pm\rho+\coth\rho\del_\pm\tau\right)\,,\\
&{\cal L}_\pm^3=\mp\frac12\text{e}^{2\tau}\l\coth\r\del_\pm\r\pm\frac12\coth^2\r(1-\l\text{e}^{2\tau})\del_\pm\tau\,,
\end{split}
\ee
where $\zeta$ is a complex parameter and we also introduce the matrix valued components
\be
{\cal L}_\pm=\sum_{a=1}^3{\cal L}_\pm^a\s_a\,,
\ee
with $\s_a$ being the Pauli matrices.  We have indeed checked that, the ${\cal L}_\pm$'s satisfy the flatness condition
\be
\del_+{\cal L}_--\del_-{\cal L}_+=[{\cal L}_+,{\cal L}_-]\,,
\ee
provided that the equations of motion of the action \eqref{action.2dHT} are satisfied.
The existence of the Lax connection ensures integrability of \eqref{action.2dHT}, where $\zeta$ plays the r\^ole of a spectral parameter, in the weak sense.
To prove integrability in the Hamiltonian sense we need to show that the conserved charges are in involution. In particular, we need to show that the spatial part of the Lax components \eqref{Lax.comp}, obeys the Maillet r/s-form~\cite{Sklyanin:1980ij,Maillet:1985ek,Maillet:1985ec}. This property is inherited from the $\l$-deformed 
$\nicefrac{SL(2,\mathbb{R})_{-k}}{SO(2)}$ $\s$-model, whose integrability in the weak and Hamiltonian sense were showed, respectively, in~\cite{Hollowood:2014rla} and~\cite{Hollowood:2015dpa}. Its connection to \eqref{action.2dHT} will be established later in the subsection~\ref{xjakqj}.

\subsection{Asymptotic limit and conformal invariance}
\label{sec.2d.conf}

Let now consider the large-$\rho$ limit of \eqref{action.2dHT} and  \eqref{2d.dilaton}, yielding
\be
\label{action.2dHT1}
\begin{split}
S=  {k\ov \pi}\int\text{d}^2\s\left(\del_+\rho\del_-\rho-\del_+\tau\del_-\tau
+\l\,\text{e}^{2\tau}\del_+(\rho+\tau)\del_-(\rho+\tau)\right)\, ,
\end{split}
\ee
with the scalar $\Phi$ being equal to
\be
\label{action.2dHT2}
\Phi=-\rho\,.
\ee
The corresponding target space geometry  (in the normalization of \eqref{gen2d}) is given by
\be
\label{metric.double}
\text{d} s^2=2k(\text{d}\rho^2-\text{d}\tau^2)+\l V\ell_\mu\ell_\nu\text{d}X^\mu\text{d}X^\nu\,,
\ee
where
\be
V=2k\text{e}^{2\tau}\,,\quad \ell_\mu\text{d}X^\mu=\text{d}\rho+\text{d}\tau\,,\quad X^\m=(\rho,\tau)
\ee
and $\ell^\m$ is a null and geodesic vector   \eqref{gen.KS.metric}. Note that the line element \eqref{metric.double}
remains invariant under translations of $\rho$ and it is flat for $\l=0$.
Let us now work out the one-loop RG flows \eqref{RG.oneloop}  for the target space metric \eqref{metric.double} and the diffeomorphism
\be
\label{qshsoq2}
\xi^\m\del_\m= g^{\m\n}\del_\n\Phi\del_\m+\frac{c}{2k}(\del_\r-\del_\tau)\quad\Longrightarrow\quad
\xi_\m\text{d}X^\m=\text{d}\widehat\Phi\,,\quad \widehat\Phi=\Phi+c(\tau+\rho)\,,
\ee
where the scalar $\Phi$ is given by \eqref{action.2dHT2}. The end result is given again by \eqref{c.dep.RG0}, where the
parameter $\l$ can be absorbed in shifts of $\tau$. Hence, we should pick $c=1$ corresponding to a {\it conformal invariant $\s$-model}
as the diffeomorphism can be written in terms of the dilaton field
$\widehat\Phi=\tau$.

\no
In what follows, we will show that this model coincides with the $\nicefrac{SL(2,\mathbb{R})_{k}}{SO(2)}$ coset CFT by rewriting \eqref{action.2dHT1}.
To do so we perform the coordinate transformation
\be
u=\text{e}^{-\rho-\tau}\,,\quad
v=\frac{1}{\l}\text{e}^{\rho-\tau}+\text{e}^{\rho+\tau}
\ee
and we find the $\nicefrac{SL(2,\mathbb{R})_k}{SO(2)}$ coset CFT in global coordinates
\be
S=\frac{k}{\pi}\int\text{d}^2\s\frac{\del_+ u\del_- v}{1- u  v}\,,\quad \Phi=-\frac12\ln(1- u v)\, .
\ee
Finally, the target space of the aforementioned model has a cosmological interpretation~\cite{Kounnas:1992wc}, as we can see using the coordinate transformation
\be
 u=\cosh t\,\text{e}^{-r}\,,\quad  v=\cosh t\,\text{e}^{r}\,,
\ee
leading to
\be
S=\frac{k}{\pi}\int\text{d}^2\s\left(-\del_+t\del_-t+\coth^2t\del_+r\del_-r\right)\,.
\ee

\subsection{Connection to the $\l$-deformed integrable models}
\label{xjakqj}

In this subsection, we will show that the action \eqref{action.2dHT} can be
derived from the $\l$-deformed model based on the $\nicefrac{SL(2,\mathbb{R})_{-k}}{SO(2)}$ exact CFT~\cite{Sfetsos:2013wia}
\begin{eqnarray}
\label{2dmetric}
 &&S=  {k\ov \pi}  \int \text{d}^2\s\,  \left\{ \frac{1 - \l}{ 1 + \l} \big(   \del_+\r\del_-\r  - \coth^2 \r \, \del_+\tau\del_-\tau \big)\right.
 \\
 &&
 +\left. \frac{4  \l}{1 - \l^2} \big(  \cosh\tau \, \del_+\r + \sinh\tau \, \coth\r \, \del_+\tau  \big)\big(  \cosh\tau \, \del_-\r + \sinh\tau \, \coth\r \, \del_-\tau   \big)\right\} \  ,\nonumber
\end{eqnarray}
and the scalar
\be
\label{2ddilaton}
\Phi=-\ln\sinh\rho\,,
\ee
 via a limiting procedure. Before that, let us first list some properties of \eqref{2dmetric}.
Inserting into \eqref{RG.oneloop}
the target space of the action \eqref{2dmetric} and the diffeomorphism $\xi_\mu=\partial_\mu\Phi$, with $\Phi$ being the dilaton \eqref{2ddilaton}, leads to the
one-loop RG flow~\cite{Itsios:2014lca}
\be
\label{RGflows.2d}
\frac{\text{d}\l}{\text{d}t}=\frac{\l}{k}\,,
\ee
where the level $k$ does not flow and the coset CFT $\nicefrac{SL(2,\mathbb{R})_{-k}}{SO(2)}$ is an IR fixed point at $\l=0$.
The action \eqref{2dmetric} describes irrelevant perturbations around the conformal point $\l=0$,  whose target space describes the
black hole of \cite{wittenbh} beyond the singularity.  This is consistent with the Weyl anomaly coefficient
\be
W=2-3(R+4\nabla^2\Phi-4(\nabla\Phi)^2)=2+\frac{6}{k}\frac{1+\lambda^2}{1-\lambda^2}\,,
\ee
matching Zamolodchikov's $C$-function~\cite{Georgiou:2018vbb,Sagkrioti:2018abh} and it satisfies the $c$-theorem~\cite{Zamolodchikov:1986gt}.
In addition, the model \eqref{2dmetric} is classically integrable as its equations of motion take the form of a Lax connection~\cite{Hollowood:2014rla}  and
it is also integrable in the Hamiltonian sense as its conserved charges are in involution~\cite{Hollowood:2015dpa}.
Finally, \eqref{2dmetric} may be rewritten in a conformally flat form in global coordinates \cite{Sfetsos:2014cea}
\be
\label{2dmetric1}
S=k\frac{1-\l^2}{\pi}\int\text{d}^2\s\frac{\del_+u\del_-v}{(u-\l v)(v-\l u)-(1-\l^2)^2}\,,
\ee
 after the transformation
\be
\label{2dmetric.coord}
u=\cosh\rho\left(\text{e}^{-\tau}+\l\text{e}^{\tau}\right)\,,\quad v=\cosh\rho\left(\text{e}^{\tau}+\l\text{e}^{-\tau}\right)\,
\ee
 and a subsequent extension of the range of the new variables $u$ and $v$.

\no
As advertized above the Kerr--Schild deformation of the coset CFT $\nicefrac{SL(2,\mathbb{R})_{-k}}{SO(2)}$  \eqref{action.2dHT} can be obtained from the
$\l$-deformed corresponding coset CFT \eqref{2dmetric} using a limiting procedure.  The appropriate limit,  asymptotic and simultaneously taking the deformation parameter to zero in a correlated way, is
\be
\label{2dmetric.coord.limit}
\tau\to\tau-\frac12\ln\varepsilon\,,\quad \l\to\varepsilon\l\quad\text{with}\quad \varepsilon\to0\, .
\ee
Equivalently, \eqref{action.2dHTconf} could be reached from \eqref{2dmetric1} in the limit
\be
\label{hdi11}
u\to\sqrt{\varepsilon\l}u\,,\quad v\to\frac{v}{\sqrt{\varepsilon\l}}\,,\quad \l\to\varepsilon\l\quad\text{with}\quad \varepsilon\to0\,.
\ee
The Lax connections of the actions \eqref{action.2dHT} and \eqref{2dmetric} coincide in the limit \eqref{2dmetric.coord.limit}.\footnote{ More precisely the Lax connection of the $\l$-deformed action \eqref{2dmetric}, as it was constructed in~\cite{Hollowood:2014rla}, turns out to depend on all three group parameters of the original $SL(2,\mathbb{R})$ element. One of them  can be absorbed in a gauge transformation of the Lax connection,
that is ${\cal L}_\pm\to h^{-1}{\cal L}_\pm h-h^{-1}\del_\pm h$ with $h\in SO(2)$. Alternatively, one may fix one of the parameters thanks to the $SO(2)$ gauge invariance.}
This property also proves the integrability of the action \eqref{action.2dHT} in the Hamiltonian sense.

\no
Finally, we would like to consider the large-$\rho$ limit in \eqref{2dmetric} and \eqref{2ddilaton} and to make contact with 
the results of~\cite{Itsios:2021wso}. Doing so we find
\begin{eqnarray}
\label{2dmetric.largerho}
 &&S=  {k\ov \pi}  \int \text{d}^2\s\,  \left\{ \frac{1 - \l}{ 1 + \l} \big(   \del_+\r\del_-\r  -  \, \del_+\tau\del_-\tau \big)\right.
 \\
 &&
 +\left. \frac{4  \l}{1 - \l^2} \big(  \cosh\tau \, \del_+\r + \sinh\tau \, \del_+\tau  \big)\big(  \cosh\tau \, \del_-\r + \sinh\tau \, \del_-\tau   \big)\right\} 
 \nonumber
\end{eqnarray}
and the scalar $\Phi=-\rho$. The latter background interpolates between a free scalar times a linear dilaton 
background in the IR and the $\text{AdS}_2$ space towards the UV~\cite{Itsios:2021wso}.\footnote{Upon taking the large-$\tau$ limit \eqref{2dmetric.coord.limit} yields again \eqref{action.2dHT1}. Hence, the large-$\rho$ and -$\tau$ limits commute.}

\no
For completeness, we also present the conformally flat form of \eqref{2dmetric.largerho}
\be
\label{sqojqoqqj}
S=k\frac{1-\l^2}{\pi}\int\text{d}^2\s\frac{\del_+u\del_-v}{(u-\l v)(v-\l u)}.
\ee
where
\be
u=\frac{\text{e}^\rho}{2}\left(\text{e}^{-\tau}+\l\text{e}^\tau\right)\,,\quad v=\frac{\text{e}^\rho}{2}\left(\text{e}^{\tau}+\l\text{e}^{-\tau}\right)\,,
\ee
which can be viewed as the large-$\rho$ limit of \eqref{2dmetric1} and \eqref{2dmetric.coord}, 
respectively.\footnote{The limit of $\l=1-\nicefrac{\k^2}{k}$
as $k\gg1$ towards the UV, can be easily taken in \eqref{sqojqoqqj} leading to the $\text{AdS}_2$ metric expressed in Poincar\'e coordinates
\begin{equation*}
S=\frac{\k^2}{2\pi}\int\text{d}^2\s\,\frac{1}{z^2}\left(\del_+z\del_-z-\del_+t\del_-t\right)\,.
\end{equation*}
A similar conclusion, with a slightly different reasoning, has been reached in \cite{Sfetsos:2014cea}.}

\begin{figure}[h!]
\begin{center}
  \includegraphics[scale=0.6]{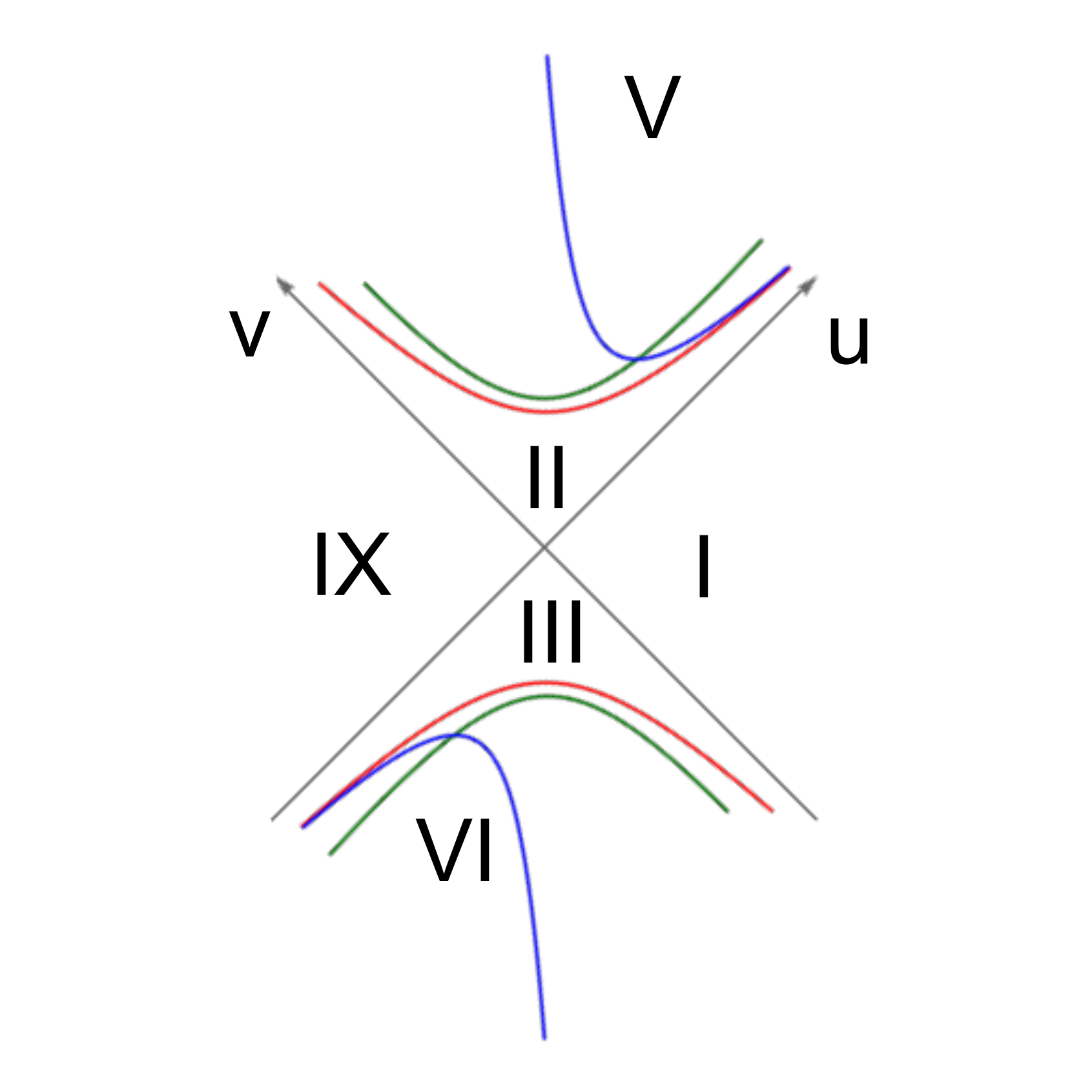}
  \end{center}
  \vskip-0.45in
  \caption{The Kruskal--Szekeres diagrams of the scale invariant \eqref{action.2dHTconf} and the $\l$-deformed model \eqref{2dmetric1} 
  (for $\l\neq 0$), showing the location of the singularity, in blue and green colors, respectively. In red is the singularity at $uv=1$ in the CFT case (for $\l=0$) of \cite{wittenbh}. The blue curve is not left-right symmetric since the limit \eqn{hdi11} breaks that symmetry. }
  \label{Kruskal.scale}
\end{figure}

\no
 We have drawn the Kruskal--Szekeres diagrams in Figure \ref{Kruskal.scale} for the actions \eqref{action.2dHTconf} and 
\eqref{2dmetric1}  (for generic $\l$ and for $\l=0$) depicting the positions of the horizons at $u=0$ and at $v=0$, as well as of the 
singularity curves corresponding to the vanishing of the functions in the denominators in the corresponding $\s$-model actions. 
Defining $v=T-X$ and $u = T+X$, in regions I-IV the time $T$
flows upwards whereas in the regions V and VI sideways. The discussion is completely analogous to that in \cite{wittenbh}.

\vskip -0.2in

\subsection{Embedding to supergravity}

The embedding to supergravity can be found by taking the asymptotic limit \eqref{2dmetric.coord.limit} in the
solution corresponding to the embedding of the $\l$-deformed $\nicefrac{SL(2,\mathbb{R})_{-k}}{SO(2)}$ non-compact CFT times
the $\l$-deformed $\nicefrac{SU(2)_{k}}{U(1)}$ compact CFT backgrounds. This was done in Section 5.3 of~\cite{Sfetsos:2014cea} and we emphasize that it is necessary to consider both $\l$-deformed spaces in order for the embedding to be achieved.
Upon applying the asymptotic limit \eqref{2dmetric.coord.limit} its NS sector is given by
\be
\small
\begin{split}
&\text{d} s^2=2k\left\{\text{d}\rho^2-\coth^2\rho\text{d}\tau^2+\text{d}\omega^2+\cot^2\omega\text{d}\phi^2+\l\text{e}^{2\tau}\,(\text{d}\rho+\coth\rho\text{d}\tau)^2\right\}
+\sum_{i=1}^6\text{d}x_i^2\,,\\
&\text{e}^{-2\Phi}=\sin^2\omega\sinh^2\rho\,.
\end{split}
\ee
These fields have to be supported by RR-flux forms in order to satisfy the type-II supergravity equations.

\no
The embedding to type-IIA supergravity requires the two- and four-forms
\be
\begin{split}
&F_2=\sqrt{2k\l}\,\text{e}^\tau(\cosh\rho\text{d}\tau+\sinh\rho\text{d}\rho)\wedge(\cos\omega\sin\phi\,\text{d}\phi+\cos\phi\sin\omega\,\text{d}\omega)\,,\\
&F_4=\sqrt{2k\l}\,\text{e}^\tau(\cosh\rho\text{d}\tau+\sinh\rho\text{d}\rho)\wedge(\cos\omega\cos\phi\,\text{d}\phi-\sin\phi\sin\omega\,\text{d}\omega)\wedge J_2\,,\\
&J_2=\text{d}x_{12}+\text{d}x_{34}+\text{d}x_{56}\,,
\end{split}
\ee
where we have used the standard notation $\text{d}x_{12}=\text{d}x_1\wedge\text{d}x_2$, etc.

\no
Similarly, the embedding to type-IIB supergravity requires the five-form
\be
\begin{split}
&F_5=(1+\ast)f_5\,,\\
&f_5=\sqrt{2k\l}\,\text{e}^\tau(\cosh\rho\text{d}\tau+\sinh\rho\text{d}\rho)\wedge(\cos\omega\sin\phi\text{d}\phi+\cos\phi\sin\omega\,\text{d}\omega)\wedge J_3\,,\\
&J_3=\text{d}x_{135}-\text{d}x_{146}-\text{d}x_{245}-\text{d}x_{236}\,.
\end{split}
\ee
These supergravity backgrounds turn out to be non-supersymmetric as one can easily see by analyzing the dilatino variation.

\section{Kerr--Schild deformation of a 3-dim target space CFT}
\label{Sec:3}

We would like to extend the results of the previous section to higher dimensional examples. However, a brut force analysis turns out to be rather tedious
and for this reason we will consider analogue asymptotic limits of $\l$-deformed orthogonal coset CFTs, along the lines of Subsection~\ref{xjakqj}.

Let us first consider the three-dimensional gauged WZW models for orthogonal groups.
The target space consists of a metric and a dilaton
given in terms of~\cite{Bars:1992ti}
\begin{equation}
\label{metogwzw3}
\text{d}s^2=2k\left(\frac{\text{d}b^2}{b^2-1}
+\frac{b-1}{b+1}
\frac{\text{d}u^2}{u(v-u-2)}
-\frac{b+1}{b-1}
\frac{\text{d}v^2}{v(v-u-2)}
\right)
\end{equation}
and
\begin{equation}
\label{dilogwzw3}
\text{e}^{-2\Phi}=
(b^2-1)(v-u-2)\,,
\end{equation}
 which are valid for large $k$.
The various cosets appear for different ranges of the global coordinates $(u,v,b)$ and are summarized in Eq. (30) of~\cite{Petropoulos:2006py}, see also below for the reader's convenience
\begin{itemize}
\item $\text{CAdS}_{3,k}=\nicefrac{SO(2,2)_{-k}}{SO(2,1)_{-k}}$
\begin{equation}
\label{ranads}
\left\{\vert b\vert>1, uv>0\right\}\quad\text{or}\quad
\left\{\vert b\vert<1, uv<0\ \text{excluding}\ 0<v<u+2<2
\right\};
\end{equation}
\item $\text{CdS}_{3,k}=\nicefrac{SO(3,1)_{k}}{SO(2,1)_{k}}$
\begin{equation}
\label{rands}
\left\{\vert b\vert<1, uv>0\right\}\quad\text{or}\quad
\left\{\vert b\vert>1, uv<0\ \text{excluding}\ 0<v<u+2<2
\right\};
\end{equation}
\item $\text{CH}_{3,k}=\nicefrac{SO(3,1)_{-k}}{SO(3)_{-k}}$
\begin{equation}
\label{ranh}
\left\{\vert b\vert>1, 0<v<u+2<2
\right\};
\end{equation}
\item $\text{CS}_{3,k}=\nicefrac{SO(4)_{k}}{SO(3)_{k}}$
\begin{equation}
\label{rans}
\left\{\vert b\vert<1, 0<v<u+2<2
\right\}.
\end{equation}
\end{itemize}
The first two have Minkowskian signature and the remaining ones Euclidean. Also, note that, none of these spaces enjoys
any Killing vector.

\subsection{The integrable $\lambda$-deformation of the $\nicefrac{SO(2,2)_{-k}}{SO(2,1)_{-k}}$ coset CFT}

The $\lambda$-deformed version of \eqref{metogwzw3} was constructed in~\cite{Demulder:2015lva}.
In our global coordinates the metric is given by
\ba
\label{lambda.metogwzw3}
&&\text{d}s_\lambda^2  =   2k  \bigg( \frac{1 + \lambda^2 + 2 \lambda (v - u - 1)}{1 - \lambda^2} \frac{\text{d}b^2}{b^2-1}
+ \frac{1 + b}{1 - b} \frac{1 + \lambda^2 + 2 \lambda (1 - v)}{1 - \lambda^2} \frac{\text{d}v^2}{v (v - u - 2)}
\nonumber
\\
&&\qquad - \frac{1 - b}{1 + b} \frac{1 + \lambda^2 + 2 \lambda (1 + u)}{1 - \lambda^2} \frac{\text{d}u^2}{u (v - u - 2)} -
\frac{4 \lambda}{1 - \lambda^2} \frac{\text{d}u \, \text{d}v}{v - u - 2}
\\
&&\qquad  - \frac{4 \lambda}{1 - \lambda^2} \frac{\text{d}b \,\text{d}u}{1 + b} - \frac{4 \lambda}{1 - \lambda^2} \frac{\text{d}b \,\text{d}v}{1 - b} \bigg)\,,
\nonumber
\ea
where the dilaton is still given by \eqref{dilogwzw3}. This model is classically integrable since the equations of motion can be written in terms of
a Lax connection~\cite{Hollowood:2014rla} and its conserved charges are in involution~\cite{Hollowood:2015dpa}.
The $\s$-model \eqref{lambda.metogwzw3} is renormalizable at one-loop order in the $\nicefrac1k$ expansion.  Indeed,
inserting \eqref{lambda.metogwzw3}, \eqref{dilogwzw3}  into \eqref{RG.oneloop}, leads to the one-loop RG flow~\cite{Sfetsos:2014jfa,Appadu:2015nfa}
 \begin{equation}
 \label{RG.coset}
 \beta^\lambda=\frac{\text{d}\l}{\text{d}t} = \frac{\l}{2 k} \, ,
\end{equation}
where the CFT point $\l=0$ is an IR fixed point. In addition, the Weyl anomaly coefficient in the case at hand reads~\cite{Georgiou:2018vbb,Sagkrioti:2018abh}
\begin{equation}
W=3-3(R+4\nabla^2\Phi-4(\nabla\Phi)^2)=3+\frac{9}{2k}\frac{1+\lambda^2}{1-\lambda^2}\,,
\end{equation}
which is identified with the Zamolodchikov's c-function~\cite{Zamolodchikov:1986gt}.

\subsection{The Kerr--Schild asymptotic limit}

\noindent
Let us now consider the metric \eqref{lambda.metogwzw3} in the parametric region of \eqref{ranads} or \eqref{rands} and take a variant of
the limit considered in~\cite{Hoare:2015wia}, followed by taking the parameter $\l$ to zero. Specifically,
\begin{equation}
\label{HTlimit.l}
(u,v)\to\frac1\varepsilon(u,v)\,,\quad \lambda\to\varepsilon\lambda\quad\text{with}\quad \varepsilon\to0\,.
\end{equation}

\noindent
Applying this limit to \eqref{lambda.metogwzw3} we find that
\begin{equation}
\label{jfjsksk}
\small
\begin{split}
&\text{d} s^2  = 2k\left(\frac{\text{d}b^2}{b^2-1}
+\frac{b-1}{b+1}
\frac{\text{d}u^2}{u(v-u)}
-\frac{b+1}{b-1}
\frac{\text{d}v^2}{v(v-u)}
\right)
+\l\,V\ell_\mu\ell_\nu\text{d}X^\mu\text{d}X^\nu\,,\\
&\ell =\text{d}\left[(b+1)v-(b-1)u\right]\,,\quad V=\frac{4\,k}{(b^2-1)(v-u)}\,,\quad X^\mu=(b,u,v)\, .
\end{split}
\end{equation}
The vector  $\ell^\mu$ is null and geodesic obeying \eqref{gen.KS.metric}. The dilaton is obtained from \eqref{dilogwzw3}
and equals to
\begin{equation}
\label{dilogwzw3HT}
\text{e}^{-2\Phi}=(b^2-1)(v-u)\, ,
\end{equation}
 where an $\varepsilon$-depended constant has been disregarded by redefining the additive constant that 
we may always add to the dilaton.
Note that the $\l$-independent term in \eqref{jfjsksk} develops an invariance involving the same
scaling of the variables $u$ and $v$, which is not present in the original metric \eqref{metogwzw3}.
This invariance is broken by the dilaton field \eqref{dilogwzw3HT},\footnote{This property was used in~\cite{Arutyunov:2015mqj},
where the authors performed an Abelian T-duality in the background \eqref{jfjsksk} (for $\l=0$), but for the range of variables corresponding to the $CH_{3,k}$ coset CFT in~\eqref{rands}, with respect to the aforementioned scaling symmetry. 
Due to the non-invariance of the dilaton the result is not expected to be and in fact it is not, a conformal background. 
Nevertheless, it turns out to be scale invariant.}
but also by the $\l$-dependent term in~\eqref{jfjsksk}.

\no
We note that this is a different limit than the one taken in~\cite{Itsios:2021wso}. In that work we considered \eqref{lambda.metogwzw3} in the parametric region of \eqref{ranh} and took the asymptotic limit $b\to\infty$ leaving the parameter $\l$ intact.
The derived $\s$-model interpolates under RG flow between the $\nicefrac{SO(3)_k}{SO(2)}$ coset CFT times a linear dilaton in the IR~\cite{Petropoulos:2006py} towards an $H_3$ hyperbolic space in the UV as $\l\to1$~\cite{Itsios:2021wso}.

\subsection{Scale invariance }

Let us now study the one-loop RG flows \eqref{RG.oneloop}  for the target
space \eqref{jfjsksk}, where we consider the diffeomorphism
\begin{equation}
\label{full.diffeomorphism}
\xi^\m\del_\m= g^{\mu\nu}\partial_\nu\Phi\,\partial_\mu+\eta\,,\quad \eta=-\frac{c}{2k}\left(u\,\partial_u+v\,\partial_v\right),
\end{equation}
with $c$ being an arbitrary parameter. Similarly to \eqref{KS.diff} the second term involves a 
diffeomorphism corresponding to the scale invariance of the $\l$-independent term in \eqn{jfjsksk}.
This $\eta$ diffeomorphism
cannot be written in terms of a new scalar and its Lie derivative is given by
\be
\label{prop2}
\nabla_\mu \eta_\nu+\nabla_\nu\eta_\mu=- \frac{2\l c}{(b^2-1)(v-u)}\ell_\mu\ell_\nu\,,
\ee
which is similar to \eqref{qshsoq11}.
Employing the one-loop RG flows \eqref{RG.oneloop} and the diffeomorphism \eqref{full.diffeomorphism}, one finds
\begin{equation}
\label{c.dep.RG}
\frac{\text{d}\lambda}{\text{d}t}=(1-c)\frac{\lambda}{2k}\,,
\end{equation}
whereas the level $k$ does not flow at this order.
Following a similar reasoning as in the two-dimensional case we choose $c=1$.
Hence, \eqn{jfjsksk} corresponds to a {\it scale invariant} $\s$-model for $\lambda$ different than zero and to a
{\it Weyl invariant} one for $\lambda$ equal to zero.

\subsection{Asymptotic limit and conformal invariance}

Let us now consider the large-$b$ limit of \eqref{jfjsksk} and \eqref{dilogwzw3HT}, yielding
\begin{equation}
\label{lambda.metogwzw3HT1}
\begin{split}
\text{d} s^2  &= 2k\left(\frac{\text{d}b^2}{b^2}
+\frac{\text{d}u^2}{u(v-u)}
-\frac{\text{d}v^2}{v(v-u)}
\right)
+\frac{4k\lambda}{b^2(v-u)}\ell_\mu\ell_\nu\text{d}X^\mu\text{d}X^\nu\, ,
\end{split}
\end{equation}
where $\ell_\mu\text{d}X^\mu=\text{d}[b(v-u)]$, with $X^\m=(b,u,v)$ and
\be
\text{e}^{-2\Phi}=b^2(v-u)\,.
\ee
The latter line element remains invariant under rescalings of $b$ as well and it is flat for $\l=0$. Let us now 
work out the one-loop RG flows \eqref{RG.oneloop} for the above metric
with the diffeomorphism
\be
\label{qshsoq3}
\xi^\m\del_\m= g^{\mu\nu}\partial_\nu\Phi\,\partial_\mu-\frac{c}{2k}\left(-b\del_b+u\,\partial_u+v\,\partial_v\right)\quad\Longrightarrow\quad
\xi_\m\text{d}X^\m=\text{d}\widehat\Phi\,,
\ee
where $\widehat\Phi=\Phi+c\ln[b(v-u)]$. The above diffeomorphism
yields the RG flow \eqref{c.dep.RG} and the parameter $\l$ can be also absorbed in (unphysical) rescalings of $(u,v)$.
Hence, we should pick $c=1$ corresponding to a {\it conformal invariant $\s$-model} since the diffeomorphism is
expressed in terms of a new dilaton field $\widehat\Phi=\nicefrac12\ln(v-u)$.

\no
We conclude by identifying the corresponding CFT. To do so, we perform the following coordinate transformation in \eqref{lambda.metogwzw3HT1}
\be
v = \text{e}^{2\tau} \cosh^2y \, ,\quad u = \text{e}^{2\tau} \sinh^2y\, ,\quad  b =\text{e}^{2\rho}\, .
\ee
The corresponding $\s$-model (in the normalization of \eqref{gen2d}) reads
\be
\label{3daction.CFT}
S=  {4k\ov \pi}\int\text{d}^2\s\left(\del_+\rho\del_-\rho-\del_+\tau\del_-\tau+\del_+y\del_-y
+4\l\,\text{e}^{2\tau}\del_+(\rho+\tau)\del_-(\rho+\tau)\right)
\ee
and the dilaton equals to $\widehat\Phi=\tau$. Following the discussion in Subsection~\ref{sec.2d.conf} we find that the model \eqref{3daction.CFT}
describes  the $\nicefrac{SL(2,\mathbb{R})_{4k}}{SO(2)}\times\mathbb{R}$ CFT.

\section{Conclusions}

In the present work we studied Kerr--Schild deformation of coset CFTs of a non-compact target spacetime. Firstly,
we considered deformations of the two-dimensional coset CFT $\nicefrac{SL(2,\mathbb{R})_{-k}}{SO(2)}$~\cite{wittenbh}. The
derived $\s$-model is classically integrable and scale rather than Weyl invariant. 
The target spacetime of this model has the interpretation of a scale invariant black hole which we embed in type-IIA and -IIB supergravity.
We show that this type of deformation can be obtained from the corresponding $\l$-deformed model by performing a particular limiting procedure.
Finally, in  an asymptotic limit the model becomes Weyl invariant, in particular the coset CFT $\nicefrac{SL(2,\mathbb{R})_k}{SO(2)}$.

\no
We extended the above results in a three-dimensional model where we considered a similar limiting procedure of the 
$\l$-deformed $\nicefrac{SO(2,2)_{-k}}{SO(2,1)_{-k}}$ coset CFT. The derived model can be viewed again as a Kerr--Schild deformation of an asymptotic limit of 
the underlying $\nicefrac{SO(2,2)_{-k}}{SO(2,1)_{-k}}$ coset CFT. As in the two-dimensional case, the model is classically integrable 
and in addition scale invariant. 
Weyl invariance is restored in an asymptotic limit, yielding the $\nicefrac{SL(2,\mathbb{R})_{4k}}{SO(2)}\times\mathbb{R}$ CFT.

\no
The results of the present work could be extended for higher dimensional symmetric coset CFTs by 
taking appropriate limiting procedures of the corresponding $\l$-deformed models. In fact we have explicitly checked the four- and five-dimensional cases, 
namely the $\l$-deformed $\nicefrac{SO(3,2)_{-k}}{SO(3,1)_{-k}}$ and $\nicefrac{SO(4,2)_{-k}}{SO(4,1)_{-k}}$
$\s$-models. 
We will not present any details of this technically involved analysis since the results are identical to the two- and three-dimensional cases.

\no
A natural question is whether scale invariance persists beyond one-loop. The simplest such exact background corresponds to the coset CFT 
$\nicefrac{SL(2,\mathbb{R})_{-k}}{SO(2)}$ and it was found in \cite{Dijkgraaf:1991ba,Bars:1992sr}.
We have studied its two-loop renormalizability~\cite{Hull:1987pc,Hull:1987yi,Metsaev:1987bc,Metsaev:1987zx,Osborn:1989bu} under Kerr--Schild deformation and indeed we have found that scale invariance persists. 
We refrain from presenting the details of the intermediate steps as they are complicated and its conceptual core remains intact.

\no
Another potential extension would be to consider Kerr--Schild deformation of Yang--Baxter symmetric cosets which were
constructed in~\cite{Delduc:2013fga,Delduc:2013qra}. These types of $\s$-models are related to the $\l$-deformed ones
by a Poisson--Lie T-duality and an analytic continuation of the coordinates and parameters of the underlying $\s$-model~\cite{Vicedo:2015pna, Hoare:2015gda,Sfetsos:2015nya,Klimcik:2015gba,Klimcik:2016rov}.

\section*{Acknowledgements}

 We would like to thank Nat Levine and Arcady Tseytlin for a useful correspondence and Andreas Stergiou for related discussions.\\
 We also thank the organizers and the participants of the conference ``Integrability, Dualities and Deformations" (Santiago de Compostela, Spain,
30 August -- 3 September 2021) for the stimulating atmosphere, where results of this work were presented by K.~Siampos.\\
 The research work of G.~Itsios  is supported by the Einstein Stiftung Berlin via the Einstein International Postdoctoral Fellowship program 
 ``Generalised dualities and their holographic applications to condensed matter physics'' (project number IPF-2020-604).
 G.~Itsios is also supported by the Deutsche Forschungsgemeinschaft (DFG, German Research Foundation) 
 via the Emmy Noether program ``Exploring the landscape of string theory flux vacua using exceptional field theory'' (project number 426510644).\\
The research work of K.~Sfetsos was supported by the Hellenic Foundation for
Research and Innovation (H.F.R.I.) under the ``First Call for H.F.R.I.
Research Projects to support Faculty members and Researchers and
the procurement of high-cost research equipment grant'' (MIS 1857, Project Number: 16519).\\
The research work of K.~Siampos has received funding from the Hellenic Foundation
for Research and Innovation (H.F.R.I.) and the General Secretariat for Research and Technology (G.S.R.T.),
under grant agreement No 15425.

\appendix

\section{Essentials of Kerr--Schild perturbations}
\label{appendix.KS.gen}

In this section we will consider metric perturbations of the Kerr--Schild form~\cite{KSpaper,GGPaper,Xanthopoulos1,Taub,Dereli:1986cm,Xanthopoulos2}
\be
\label{metric.KS}
 g_{\mu\nu}=g^{(0)}_{\mu\nu}+\l\,h_{\mu\nu}\,,\quad h_{\mu\nu}=V\ell_\mu\ell_\nu\,,\quad \mu=1,2,\dots,D\,,
\ee
where $\l$ is a booking keeping (expansion) parameter and $V$ is the Kerr--Schild potential.
In addition, the vector $\ell^\mu$ is null and geodesic
\be
\label{condi.KS}
g^{(0)}_{\mu\nu}\ell^\mu\ell^\nu=0=g_{\mu\nu}\ell^\mu\ell^\nu\,,\quad \ell^\mu\nabla^{(0)}_\mu\ell_\nu=0\, .
\ee
From the above it can be easily seen that the inverse of the metric \eqref{metric.KS} is given by
\be
g^{\mu\nu}=g_{(0)}^{\mu\nu}-\l h^{\mu\nu}\,,\quad h^{\mu\nu}=V\ell^\mu\ell^\nu\,.
\ee
Using the latter and \eqref{condi.KS}, we find the Christoffel symbols
\be
\label{Christ.KS}
\G_{\k\l}{}^\m=\G^{(0)}_{\k\l}{}^\m+S_{\k\l}{}^\m\,,
\ee
where
\be
\label{Christ.KSb}
\begin{split}
S_{\k\l}{}^\m&=\l S^{(1)}_{\k\l}{}^\m+\l^2S^{(2)}_{\k\l}{}^\m\,,\\
S^{(1)}_{\k\l}{}^\m&=\frac12g^{\mu\rho}\left(\nabla^{(0)}_\k h_{\l\rho}+\nabla^{(0)}_\l h_{\k\rho}-\nabla^{(0)}_\rho h_{\k\l}\right)\, ,
\\
S^{(2)}_{\k\l}{}^\m&=\frac{1}{2}h^{\m\r}\nabla^{(0)}_\r h_{\k\l}=\frac{1}{2}h^{\m\r}h_{\k\l}\del_\r\ln V\, .
\end{split}
\ee
Using the above one may easily demonstrate that the geodesic property of the vector holds for the metric $g_{\m\n}$ as well, i.e.
\be
\ell^\mu\nabla_\mu\ell_\nu =0 \,.
\ee
We also note the property
\be
S^{(i)}_{\k\l}{}^\l=0\,,\quad i=1,2.
\ee
Next we evaluate the Riemann tensor, which takes the form
\be
R^\mu{}_{\rho\k\l}=R^{(0)\mu}{}_{\r\k\l}+\nabla^{(0)}_\k S_{\l\r}{}^\m-\nabla^{(0)}_\l S_{\k\r}{}^\m+S_{\k\s}{}^\m S_{\l\r}{}^\s-S_{\l\s}{}^\m S_{\k\r}{}^\s\, ,
\ee
implying that the corresponding Ricci tensor could be at most fourth order in $\l$. Using
\eqref{condi.KS}, we find after some algebra that the Ricci tensor truncates at second order in $\l$ and it is given by
\be
\label{kaqoqo}
R_{\m\n}=R^{(0)}_{\m\n}+\l R^{(1)}_{\mu\nu}+\l^2R^{(2)}_{\mu\nu}\,,
\ee
where
\be
\label{kaqoqob}
\begin{split}
R_{\mu\nu}^{(1)}&=\frac12\nabla^{(0)\rho}\left(\nabla^{(0)}_\m h_{\n\r}+\nabla^{(0)}_\n h_{\m\r}-\nabla^{(0)}_\r h_{\m\n}\right)\,,\\
R_{\mu\nu}^{(2)}&=\frac{1}{2}\left(\nabla^{(0)}_k(\ell^k\ell^\l\del_\l V)+V\,\big(\nabla^{(0)}_\k  \ell_\l\big)(\nabla^{(0)\k}\ell^\l-\nabla^{(0)\l}\ell^\k)\right)h_{\mu\nu}\,.
\end{split}
\ee
The above can be rewritten as\footnote{Where we have used the identity
$$
V\left(\nabla^{(0)}_\k(\ell^\k\ell^\l\del_\l V)+V(\nabla^{(0)}_{\k}\ell_{\s}-\nabla^{(0)}_{\k}\ell_{\s})\nabla^{(0)\s}\ell^\k\right)
=h_\m{}^\s\nabla^{(0)\rho}\left(\nabla^{(0)}_\n h_{\s\r}+\nabla^{(0)}_\s h_{\r\n}-\nabla^{(0)}_\r h_{\s\n}\right)\,.
$$}
\be
\begin{split}
R_{\m\n}&=R^{(0)}_{\m\n}+\frac\l2\nabla^{(0)\rho}\left(\nabla^{(0)}_\m h_{\n\r}+\nabla^{(0)}_\n h_{\m\r}-\nabla^{(0)}_\r h_{\m\n}\right)\\
&+\frac{\l^2}{2}h_\m{}^\s\nabla^{(0)\rho}\left(\nabla^{(0)}_\n h_{\s\r}+\nabla^{(0)}_\s h_{\r\n}-\nabla^{(0)}_\r h_{\s\n}\right)\,.
\end{split}
\ee
or in the suggestive form
\be
\label{sdfkskks}
R^\m{}_\n=R^{(0)\m}{}_\n-\l h^\m{}_\r R^{(0)\r}{}_\n+\frac\l2\nabla^{(0)}_\r\left(\nabla^{(0)}_\n h^{\m\r}+\nabla^{(0)\m} h_\n{}^\r-\nabla^{(0)\r} h^\m{}_\n\right)\,.
\ee
This unveils that in this case the {\it linearized approximation} is also an exact,\footnote{This property is realized in the mixed components form \eqref{sdfkskks}; with one index up and one down.}  where the metric $g_{\mu\nu}$ is given 
by \eqref{metric.KS} and $\ell^\mu$ satisfies \eqref{condi.KS}.
 This has been observed in 
\cite{Dereli:1986cm}.
Finally, we can also compute the Ricci tensor
\be
R=R^{(0)}-\l h^{\mu\nu}R^{(0)}_{\mu\nu}+\l\nabla^{(0)}_\mu\left(\nabla^{(0)}_\nu\left(V\ell^\nu\right)\ell^\mu\right)\,.
\ee
As a non-trivial example of the Kerr--Schild formulation~\cite{KSpaper} we present the renowned Kerr solution of mass $M$ and rotation parameter $a$, in the presence of a cosmological constant $\L$.  The first step is to rewrite the (A)dS metric, which will be the seed solution, in a rotating form using spheroidal coordinates\cite{Gibbons:2004uw} and in the conventions of \cite{Carrillo-Gonzalez:2017iyj}
\be
\begin{split}
g^{(0)}_{\m\n}\text{d}x^\m\text{d}x^\n&=-\frac{\D}{\Om}\left(1+\frac{\L r^2}{3}\right)\text{d}t^2+\frac{r^2+a^2\cos^2\th}{r^2+a^2}\frac{\text{d}r^2}{1+\frac{\L r^2}{3}}\\
&+\frac{r^2+a^2\cos^2\th}{\D}\text{d}\th^2+\frac{(r^2+a^2)\sin^2\th}{\Om}\text{d}\phi^2\,,
\end{split}
\ee
where
\be
\D=1-\frac{\L}{3}a^2\cos^2\th\,,\quad \Om=1-\frac{\L}{3}a^2\,.
\ee
It turns out that the Kerr--Schild potential $V$ and the null and geodesic vector $\ell^\mu$ read
\be
V=\frac{2GM r}{r^2+a^2\cos^2\th}\,,\quad 
\ell=-\frac{\D}{\Om}\text{d}t+\frac{r^2+a^2\cos^2\th}{r^2+a^2}\frac{\text{d}r}{1+\frac{\L r^2}{3}}-\frac{a\sin^2\th}{\Om}\text{d}\phi\,,
\ee
such that $R_{\m\n}=\L g_{\m\n}$. 

\no
 Further investigations and examples concerning Kerr--Schild perturbations can found in \cite{Coll:2000rm, Malek:2010mh, Malek:2011zz}.

\end{document}